\font\tenrm=cmr10

\font\ti=cmti12

\def\pmb#1{\setbox0=\hbox{#1}%
  \kern-.025em\copy0\kern-\wd0
  \kern.05em\copy0\kern-\wd0
  \kern-.025em\raise.0433em\box0}

\def\bfxi{\pmb{$\xi$}}
\documentstyle[12pt]{article}
\begin{document}
\begin{titlepage}
\begin{flushright}
IFUP-TH-49/93
\end{flushright}
\vskip .9truecm
\begin{center}
\LARGE \bf
STATIONARY SOLUTIONS IN 2+1 DIMENSIONAL GRAVITY\\
AND CLOSED TIME-LIKE CURVES\footnote{Work
partially supported by M.U.R.S.T}
\end{center}
\vskip 1truecm
\begin{center}
{{P. Menotti} \\ %[-4pt]
{\ti Dipartimento di Fisica dell' Universit\`a, Pisa 56100,
Italy and}\\ %[-5.5pt]
{\ti INFN, Sezione di Pisa}\\%[+5pt]
{\tenrm and}\\%[+5pt]
{  D. Seminara} \\ %[-4pt]
{\ti Scuola Normale Superiore, Pisa 56100, Italy and}\\ %[-5.5pt]
{\ti INFN, Sezione di Pisa}}
\end{center}
%
%\begin{document}
%\maketitle
%
\vskip .8truecm
\begin{center}
Talk presented at Europhysics Conference on High Energy Physics,\\
Marseille, France, July 22-28, 1993
\end{center}
\end{titlepage}
\begin{abstract}
We apply the reduced radial gauge to give the general solution of the
metric in 2+1 dimensions in term of quadratures. It allows a complete
controll on the support of the source. We use the result to prove that for a
general stationary universe, conical at infinity, the weak energy condition
and the absence of CTC at space infinity prevent the occurrrence of any CTC.
\end{abstract}

\vskip 5mm
Recently there has been a lot of interest, triggered by the work of
Gott \cite{Gott}, on the possible occurrence of closed time-like curves in 2+1
dimensions . However Carrol et al. \cite{Gott} proved that an open universe
with total time like momentum and buildt up of point-like spinless
particles cannot contain Gott's pairs. For a closed universe of point-like
spinless particles 't Hooft \cite{Gott} proved that if two particles meet
satisfying Gott's condition, the universe ends in a crunch, thus
suppressing the formation of CTC. On the other hand for a point
particle (or straight cosmic string) with angular momentum $J$ one has
the metric \cite{classol}
\begin{equation}
ds^2=(dt+ 4 G J dt)^2- (1-\frac{\beta}{2\pi})^2 \rho^2 d\theta^2 -d\rho^2
\end{equation}
which for $\rho^2<(4GJ)^2(1-\beta/2 \pi)^{-2}$ has CTC. Such an
uncomfortable issue is usually blamed on the singular nature of point
like sources with non zero angular momentum. Thus it is of interest to
examine what happens with sources which are extended and not
necessarily of powder like nature. The reduced radial gauge, defined
by\cite{MS1}
\begin{equation}
\xi^i\Gamma^a_{bi}(\xi)=0 \ \ {\rm and} \ \
\xi^i e^a_{i}(\xi)=\xi^i\delta^a_i,
\end{equation}
proves particularly powerful to solve the stationary problem at hand,
both in presence and in absence of rotational symmetry.
The great advantage of
the reduced radial gauge, which in our instance is a special case of the
Fermi-Walker coordinate frame in the setting of the first order
formalism, is that   one has quadrature formulas expressing the metric
in terms of the Riemann tensor, which in 2+1 dimensions practically
identifies with the energy momentum tensor $\tau^{ac}$. However
$\tau^{ac}$ is not arbitrary but is subject to the covariant
conservation and symmetry constraints
\begin{equation}
{\cal D} T^a=0\ \ {\rm and}\ \
\varepsilon_{abc} T^b\wedge e^c =0.
\end{equation}
In this formalism the metric becomes
\begin{eqnarray}
&&ds^2= (A_1^2-B_1^2) dt^2 +
2 (A_1 A_2-B_1 B_2)dtd\theta +\nonumber\\
&&(A_2^2-B^2_2) d\theta^2-d\rho^2,
\end{eqnarray}
where $A_1+1$, $A_2$, $B_1$, $B_2$ are the primitive of the
fundamental functions $\alpha_i$, $\beta_i$ which express the energy
momentum tensor in terms of the cotangent vectors
$\displaystyle{T_{\mu}=\frac{\partial \xi^0}{\partial \xi^\mu}}$,
$\displaystyle{P_{\mu}=\frac{\partial \rho}{\partial \xi^\mu}}$ and
$\displaystyle{\Theta_{\mu}=\rho\frac{\partial \theta}{\partial \xi^\mu}}$
of the
polar coordinates as follows;
\begin{eqnarray}
\label{tab}
&&\hskip-18pt\tau^{ca}(\bfxi)=e^a_\mu (\bfxi)\tau^{\mu c}(\bfxi)=\nonumber\\
&&\hskip -18pt-\frac{1}{\kappa \rho}\biggl \{
T^a \biggl [ T^c \biggl (A_2 \beta^\prime_1-A_1 \beta^\prime_2 \biggr
)+\Theta^c \biggl ( A_2 \alpha_1^\prime-A_1\alpha_2^\prime\biggr
)\nonumber\\
&&\hskip-18pt
+P^c\biggl (A_2 \gamma_1^\prime-A_1\gamma_2^\prime\biggr )\biggr ]
+\Theta^a \biggl [ T^c \biggl (B_2 \beta^\prime_1-B_1 \beta^\prime_2
\biggr )\nonumber\\
&&\hskip-18pt +\Theta^c \biggl ( B_2
\alpha_1^\prime-B_1\alpha_2^\prime\biggr )
+P^c\biggl (B_2 \gamma_1^\prime-B_1\gamma_2^\prime\biggr )\biggr ]\nonumber\\
&&\hskip -18pt +P^a \biggl [ T^c \biggl (\alpha_1 \gamma_2-\alpha_2
\gamma_1+\frac{\partial
\beta_1}{\partial \theta} \biggr )+
\Theta^c \biggl (\beta_1 \gamma_2-\beta_2 \gamma_1\nonumber\\
&&\hskip-18pt+\frac{\partial\alpha_1}{\partial \theta}  \biggr )+
P^c\biggl (\alpha_1 \beta_2-\alpha_2 \beta_1+\frac{\partial
\gamma_1}{\partial \theta} \biggr )\biggr ]\biggr \}
\end{eqnarray}
The conservation equation is automatically satisfied \cite{MS3}
by (\ref{tab}), while the symmetry condition becomes
\begin{eqnarray}
\label{symmetry1}
&&A_1\alpha_2-A_2 \alpha_1-B_1\beta_2+B_2
\beta_1=0\\
\label{symmetry2}
&&A_2\gamma_1 -A_1
\gamma_2 +\frac{\partial B_1}{\partial\theta}=0\\
\label{symmetry3}
&&B_2\gamma_1-B_1
\gamma_2 +\frac{\partial A_1}{\partial\theta}=0.
\end{eqnarray}
The described gauge allows a complete control on the support of the energy
momentum tensor in the general case \cite{MS3}.
For
conciseness sake we shall refer from now on to the rotationally invariant case
even if the argument can be carried through with an almost identical
procedure  in the general stationary case \cite{MS3}.
In the simplified situation
with rotational symmetry we have $\gamma_i=0$  and the support conditions
become
$\alpha^\prime_i=\beta^\prime_i=0$ and $\alpha_1\beta_2-\alpha_2\beta_1=0$
outside the source i.e. for $\rho>\rho_0$
It is immediately  seen that a CTC implies the existence of a CTC with
constant $\rho$ and $t$.
 Thus to prove that CTC cannot exist
is sufficient to prove that $g_{\theta\theta}$ which by assumption is
negative at space
infinity, cannot change sign. It is possible to prove that
if the determinant of the dreibein in the
radial gauge,  where
$g_{\rho\rho}\equiv
-1$, vanishes for a certain ${\bar{\rho}}$,
either the geometry is singular or the manifold becomes singular or the
universe closes at $\rho=\bar\rho$.
Thus for an open universe we have that $\det(e)>0$ for
$\rho>0$;
in particular if we denote with $\rho_0$ a point outside the source we must
have $\displaystyle{
 \frac{d {\rm det}(e)}{d\rho}\mid_{\rho=\rho_0}\ge 0}$
because  for $\rho\ge \rho_0$, due to the support conditions, the determinant
is a linear  function of $\rho$.
We show now that the WEC combined with $\det(e)>0$ and the absence
of CTC at infinity imply the absence of CTC for any $\rho$. In fact
the WEC  $v_a \tau^{ab} v_b\ge 0$, for $v^a v_a\ge 0$,
applied to the vectors $(1,1,0)$ and
$(1,-1,0)$ gives
\begin{equation}
\frac{d}{d\rho} \left [ (\alpha_2\pm\beta_2)(B_1\pm A_1)-
                 (B_2\pm A_2)(\alpha_1\pm\beta_1)\right ]\ge 0
\end{equation}
which can be integrated between $\rho$ and any point $\rho_0$ outside
the source to give
\begin{eqnarray}
\label{WECdelta}
&&\hskip-18ptE^{(\pm)} (\rho)\equiv
(B_2\pm A_2)(\alpha_1\pm\beta_1)-
(\alpha_2\pm\beta_2)(B_1\pm A_1)\ge\nonumber\\
&&\hskip-18pt(B^0_2\pm A^0_2)(\alpha^0_1\pm\beta^0_1)-
(\alpha^0_2\pm\beta^0_2)(B^0_1\pm A^0_1)
\end{eqnarray}
For a conical universe in absence of CTC at infinity we have
\begin{equation}
\label{B}
(\alpha^0_2)^2-(\beta^0_2)^2<0.
\end{equation}
 Using the symmetry equation (\ref{symmetry1}) and the
support equation (\ref{tab}) the r.h.s. of (\ref{WECdelta}) becomes
\begin{equation}
E^{(\pm)}(\rho_0)=-\frac{\alpha^0_2\pm\beta^0_2}{\alpha^0_2\mp\beta_2^0}
\frac{d}{d\rho} {\rm det}(e)\mid_{\rho=\rho_0}
\end{equation}
which due to (\ref{B}) and
$\displaystyle{\frac{d {\rm det}(e)}{d\rho}\mid_{\rho=\rho_0}\ge 0}$
is non negative i.e. $E^{(\pm)}(\rho_0)\ge 0$.\\
Let us now consider the following combination
\begin{equation}
(A_2-B_2)^2 E^{(+)}(\rho)+(A_2+B_2)^2 E^{(-)}(\rho)\ge 0.
\end{equation}
A little algebra shows that the l.h.s. equals
\begin{equation}
-2 {\rm det}^2(e) \frac{d}{d\rho} \left (\frac{g_{\theta\theta}(\rho)}{{\rm
det}(e)}\right ).
\end{equation}
Thus we reached the conclusion that
\begin{equation}
\displaystyle{\frac{d}{d\rho}
\left (\frac{ g_{\theta\theta}(\rho)}{{\rm det}(e)}\right )}\le 0;
\end{equation}
which means  that
$\displaystyle{
\left (\frac{ g_{\theta\theta}(\rho)}{{\rm det}(e)}\right )}$ is
a non increasing function
of $\rho$. As $\displaystyle{
\left (
\frac{ g_{\theta\theta}(\rho)}{{\rm det}(e)}\right )}$  at the origin is
zero we  obtain
that $g_{\theta\theta}=A^2_2(\rho)-
B^2_2(\rho)$ is always negative and thus we cannot have CTC.\\
This reasoning can be extended to all universes
except for the peculiar case of the cylindrical universe generated
by a string with tension and total angular momentum zero, even if we have
no example of extended sources satisfying the energy condition and
producing such a universe with CTC.
It is not difficult to produce examples in which all energy conditions are
satisfied
and CTC exist for any radius larger than a given one; but we obviously violate
the
condition that no CTC exists at infinity. Also renouncing to the energy
conditions
one can produce examples \cite{MS1} in which non CTC exist at infinity,
but they
exist at
finite radius.
The treatment can be readly extended with the same results to the general
stationary problem, without rotational symmetry provided universe is
conical at space infinity.

We mention that the approach with the radial gauge,
for which the solution can be given in terms
of quadratures \cite{MS1,MS2}, can be developped
also for the time-dependent situation \cite{MS2}.
% works also for the case of
% rotational symmetry with time dependent sources\cite{MS2}.

\end{document}